\newfont{\eufm}{eufm10 scaled\magstep1} 
\newfont{\msym}{msym10 scaled\magstep1 }  
\newfont{\rsym}{msym10 scaled \magstep4}  

\newcommand{\bfsp}[1]{\mbox{\msym #1\/}}

\newfont{\Bold}{cmbx10 scaled\magstep2}
\newfont{\BOLD}{cmbx10 scaled\magstep3}

\def\div{\mathop{\rm div}}
\def\curl{\mathop{\rm curl}}

\documentstyle[12pt]{siam}
\begin{document}
\title{
periodic solutions of a system of \\coupled oscillators near resonance
\thanks{This document was written \ \today .
This research was supported by
the Air Force office of Scientific Research and the National Science
Foundation under the
grant  DMS-9022621.
}}
\author{
Carmen Chicone
\thanks{Department of Mathematics, University of Missouri,
Columbia, MO 65211.}
}
\maketitle
\begin{abstract}
A system of autonomous ordinary differential
equations depending on a small parameter is considered such that the
unperturbed system has an invariant manifold of
periodic solutions that is not normally hyperbolic but
is normally nondegenerate. The bifurcation function whose zeros
are the bifurcation points for families of perturbed periodic solutions
is determined. This result is applied to find the periodic
solutions near resonance for a two degree of freedom mechanical system
modeling a rotor interacting with an elastic support.
\end{abstract}
\begin{AMSMOS} 58F14, 58F22, 58F30, 34C15 34C25.
\end{AMSMOS}
\section{Introduction}
In this paper we describe an application of the results in \cite{cccc}
to the bifurcation of periodic solutions in a smooth system of coupled
oscillators $E_\epsilon$ given by
\begin{eqnarray*}
\dot x_1&=&f_1(x_1)+\epsilon g_1(x_1,\dot x_1,x_2,\dot x_2), \\
\dot x_2&=&f_2(x_2)+\epsilon g_2(x_1,\dot x_1,x_2,\dot x_2) \
\end{eqnarray*}
where $x_i\in {\bfsp R}^2,$ $i=1,2$ and $\epsilon\in {\bfsp R}$ when the
unperturbed
system $E_0$ satisfies the following conditions:
\begin{enumerate}
\item[1.] The plane autonomous system $\dot x_1=f_1(x_1)$ has an invariant
annulus
$A$ consisting of periodic solutions (a period annulus) and every periodic
solution
in $A$ has the same period, $\eta_1>0.$ Such a period annulus is called
isochronous with period $\eta.$
\item[2.] The plane autonomous system $\dot x_2=f_2(x_2)$ has a periodic
trajectory
$\Gamma$ with period $\eta_2>0$ such that either $\Gamma$ is a hyperbolic limit
cycle or $\Gamma$ belongs to a period annulus and the derivative of an
associated
period function at $\Gamma$ does not vanish.
\item[3.] There are relatively prime positive integers $K_1$ and $K_2$ such
that $K_1\eta_1=K_2\eta_2.$ In this case we say the
periodic trajectory $\Gamma$ is in resonance with
the period annulus $A.$
\end{enumerate}

A few comments are in order on the conditions just stated. The prime example of
an isochronous period annulus is a period annulus of a linear system.
However, given any period annulus and any Poincar\'e section at a point in the
period
annulus, there is an associated period function that assigns to each point on
the
section the time of first return to the section. It is easy to see the
requirement
of a nonzero derivative of a period function as in $(2)$ above is independent
of the choice of section and the point chosen on the periodic trajectory.
The hypotheses ensure that $A\times \Gamma$ is an invariant submanifold of
the state space for the unperturbed system $E_0$ of a special type we call
a normally nondegenerate period manifold.
The condition of normal nondegeneracy defined precisely in \S 2
ensures the first order bifurcation theory in \cite{cccc} can be applied and
the existence of periodic solutions for the perturbed system near the period
manifold can be generically determined by computing the simple zeros of a
certain bifurcation function also defined in \S 2. Of particular interest
here is the fact
that the period manifold for $E_0$ is not normally hyperbolic. Thus, while
the period manifold usually does not persist after perturbation, some of
the periodic solutions on the period manifold can persist. The bifurcation
function determines the number and the position of these persistent
periodic solutions. In this way entrainment phenomena can be studied for
perturbations of systems which do not already contain stable periodic
solutions.
For background material on bifurcation problems of this type in addition
to \cite{cccfe}, \cite{cccc} the following
references and their bibliographies are suggested
\cite{avk}, \cite{arnold}, \cite{arnold1}, \cite{ch}, \cite{gh},
\cite{hayashi}, \cite{melnikov}, \cite{minorsky},
\cite{neishtadt}, \cite{rh}, \cite{rkm}, \cite{sv}, \cite{stoker},
\cite{wiggins1}, \cite{wiggins2}.

While higher dimensional  systems can be studied by the same methods, the
four dimensional system
$E_\epsilon$  illustrates the important features of the general theory and
is  sufficiently general to have many interesting specializations to physical
applications. In \S 3 we apply the theory to an ubiquitous system of
differential equations which we interpret, as in \cite{sv},
as a model for a rotor interacting
with an elastic support. We show the existence of a normally nondegenerate
period manifold in case the unperturbed system is weakly nonlinear and
also in the fully nonlinear case which corresponds to the rotor strongly
influenced by a gravitational field.
In both cases the bifurcation function is computed explicitly and
the existence of periodic solutions relative to the choice of
parameters is  determined. These results are augmented by some numerical
evidence suggesting the role of these bifurcating families of periodic
solutions in determining the global
behavior of the perturbed system.

The plan of the paper is as follows. In \S 2 we review the general theory
of \cite{cccc}. In \S 3 we specialize the general theory to the case
represented by $E_\epsilon$ and identify  the bifurcation function.
These results are applied in \S 4 to the mechanical system modeling the
rotor with elastic support. There  the bifurcation
function is computed explicitly in terms of elliptic functions  and its
zeros are computed. This determines the perturbed periodic
solutions of the coupled mechanical oscillators near resonance.
In addition, \S 4 contains a discussion of some numerical experiments
that suggest the coexistence, for certain choices of the parameters,
of perturbed periodic attractors, as predicted
by the bifurcation analysis, and more complicated nonperiodic attractors.
\section{Bifurcation Theory}
In this section we outline for completeness a result in \cite{cccc}
which will be used in the analysis of the system $E_\epsilon$ defined in the
introduction.
The analysis begins with a smooth system of differential
equations $F_\epsilon$ given by
\[
\dot x=f(x)+\epsilon g(x,\dot x,\epsilon),
\;\;x\in {\bfsp R}^{n+1},\;\;\epsilon\in {\bfsp R}\]
where the unperturbed system $F_0$ contains a normally nondegenerate period
manifold.
Here, a period manifold ${\cal A}$ is a smooth invariant connected $(k+1)$-
dimensional submanifold of ${\bfsp R}^{n+1}$
consisting entirely of periodic solutions of the unperturbed system with the
additional property that the Poincar\'e map $P$ associated with any Poincar\'e
section $\Sigma$ is the identity on ${\cal A}\cap \Sigma.$
Of course,  period manifolds generalize to many dimensions  the
concept of a period annulus.
To define the concept of normal nondegeneracy we need a few more definitions.
Restricting to a particular Poincar\'e section $\Sigma_0$ which has nonempty
intersection with ${\cal A},$
there is some $\epsilon_0>0$ and some subsection $\Sigma\subseteq \Sigma_0$
such that the parametrized Poincar\'e map
$P:\Sigma\times (-\epsilon_0,\epsilon_0)\to \Sigma_0$ given by
$(\xi,\epsilon)\mapsto P(\xi,\epsilon)$ where $P(\xi,\epsilon)$ denotes the
first
return to $\Sigma_0$ of the perturbed solution starting at $\xi\in \Sigma.$
After choosing coordinates on $\Sigma,$ given by $s: {\bfsp R}^n\to \Sigma,$
the parameterized Poincar\'e map is identified
with its local representation
$p:{\bfsp R}^n\times (-\epsilon_0,\epsilon_0)\to {\bfsp R}^n$
given by $p(y,\epsilon):=s^{-1}P(s(y),\epsilon)$  This, in turn, allows us to
define
the parametrized displacement function
$\delta:{\bfsp R}^n\times (-\epsilon_0,\epsilon_0)\to {\bfsp R}^n$ by
$\delta(y,\epsilon):=p(y,\epsilon)-y.$ Now, for $y_*\in {\bfsp R}^n$
such that $s(y_*)\in \Sigma\cap {\cal A},$
it is clear that the derivative of the map $y\mapsto \delta(y,0),$
which we denote by $D\delta(y,0),$
when evaluated at $y_*$ will have a nontrivial kernel corresponding to the
tangent
space of ${\cal A}.$ More precisely, if $v\in {\bfsp R}^n$
and
$Ds(y_*)v\in T_{s(y_*)}\Sigma\cap T_{s(y_*)}{\cal A},$
then $D\delta(y_*,0)v=0.$ Since $\Sigma\cap {\cal A}$ is $k$-dimensional,
the kernel of  $D\delta(y_*,0)$ has dimension at least $k.$ If this kernel
has dimension $k$ for each $y$ such that $s(y)\in {\cal A},$ we say
${\cal A}$ is normally nondegenerate. Perhaps a remark is in order on the
definition
of displacement. One must exercise caution when defining displacement on the
manifold $\Sigma.$ We have avoided the differential geometry necessary to give
an intrinsic definition by introducing local coordinates. However, it should be
clear
that the zero set of the displacement function, the set corresponding
to periodic solutions of $F_\epsilon,$ is invariant under change of
coordinates.

A goal of the theory in \cite{cccc} is the identification of a bifurcation
function ${\cal B}$ defined on  $\Sigma\cap {\cal A}$ whose simple zeros
correspond
to the initial values of persistent periodic solutions of the unperturbed
system.
To construct the bifurcation function, we start with a splitting of the
tangent bundle over ${\bfsp R}^{n+1}$
into three subbundles, ${\cal E}$ generated by the unperturbed vector field,
${\cal E}^{\mbox{tan}}$ tangent to ${\cal A}$ but complementary to ${\cal E},$
and ${\cal E}^{\mbox{nor}}$ normal to ${\cal A}.$ In particular, for
$y\in {\cal A}$ we have
$
{\bfsp R}^{n+1}={\cal E}(y)\oplus{\cal E}^{\mbox{tan}}(y)\oplus
{\cal E}^{\mbox{nor}}(y).
$
Such a splitting always exists but the last two summands are not unique.
Next, we define  special coordinates on  ${\bfsp R}^{n+1}$ near each point
$y\in \Sigma\times {\cal A}$ which respect the splitting.
For this, we choose
$\Delta:{\bfsp R}\times {\bfsp R}^k\times {\bfsp R}^{n-k}\to {\bfsp R}^{n+1}$
given
by $(s,\theta,\zeta)\mapsto\Delta(s,\theta,\zeta)$ such that (using
subscripted variables to denote partial derivatives)
\begin{eqnarray*}
\Delta_s(0,\theta,0):{\bfsp R}&\to&{\cal E}(\Delta(0,\theta,0)),  \\
\Delta_\theta(0,\theta,0):{\bfsp R}^k&\to&{\cal
E}^{\mbox{tan}}(\Delta(0,\theta,0)), \\
\Delta_\zeta(0,\theta,0):
{\bfsp R}^{n-k}&\to&{\cal E}^{\mbox{nor}}(\Delta(0,\theta,0)). \\
\end{eqnarray*}
Such coordinates are called adapted to the splitting over ${\cal A}.$
An associated Poincar\'e section, again denoted by $\Sigma,$
is given by the image of the map
$(\theta,\zeta)\mapsto \Delta(0,\theta,\zeta).$
In these coordinated the kernel of $D\delta(\Delta(0,\theta,0),0)$
corresponds to ${\cal E}^{\mbox{tan}}(\Delta(0,\theta,0))$
and there is a $k$-dimensional complement to the range of this derivative in
${\bfsp R}^{n+1}.$ After choosing coordinates on the range,
the linear projection $H(\theta)$ from the tangent space of ${\bfsp R}^{n+1}$
to this range can be represented as a linear map of the
form
\[
H(\theta):{\cal E}\oplus{\cal E}^{\mbox{tan}}\oplus
{\cal E}^{\mbox{nor}}(\Delta(0,\theta,0))\to {\bfsp R}^k.
\]
Next, let $t\mapsto x(t,\theta)$ denote the  solution of $F_0$ with
initial condition $x(0,\theta)=\Delta(0,\theta,0)$ and consider the variational
equation along this solution, namely,
\[\dot W=Df(x(t,\theta))W.\]
This variational equation  has a fundamental matrix solution
$t\mapsto\Phi(t,\theta)$ with initial
value $\Phi(0,\theta)=I.$
There are parametrized linear maps
\[
\begin{array}{ll}
a(t,\theta):{\cal E}^{\mbox{nor}}(x(0,\theta))\to
{\cal E}^{\mbox{tan}}(x(t,\theta)),&
b(t,\theta):{\cal E}^{\mbox{nor}}(x(0,\theta))\to
{\cal E}^{\mbox{nor}}(x(t,\theta)),\\
c(t,\theta):{\cal E}^{\mbox{tan}}(x(0,\theta))\to
{\cal E}^{\mbox{tan}}(x(t,\theta)),&
d(t,\theta):{\cal E}^{\mbox{nor}}(x(0,\theta))\to
{\cal E}(x(t,\theta)), \\
e(t,\theta):{\cal E}^{\mbox{tan}}(x(0,\theta))\to{\cal E}(x(t,\theta)),&
\end{array}
\]
such that the block form of $\Phi(t,\theta)$ with respect to the splitting is
\[
\Phi(t,\theta)=\left(
\begin{array}{ccc}
1&e(t,\theta)&d(t,\theta)  \\
0&c(t,\theta)&a(t,\theta)  \\
0&0&b(t,\theta)
\end{array}
\right)
\]
and such that
\[
e(0,\theta)=0,\;\;d(0,\theta)=0,\;\;c(0,\theta)=I,\;\;a(0,\theta)=0,\;\;b(0,\theta)=I.
\]
Also, the vector field along the unperturbed solution defined by the
perturbation,
namely,
$G(t,\theta):=g(x(t,\theta),\dot x(t,\theta),0)),$ has a representation
relative to the splitting  given by
\[
G(t,\theta)=
\left(
\begin{array}{l}
G^{\cal E}(t,\theta)\\
G^{\mbox{tan}}(t,\theta)\\
G^{\mbox{nor}}(t,\theta)
\end{array}
\right).
\]
Here, $G(t,\theta)$ is the derivative of  $f(x)+\epsilon g(x,\dot x,\epsilon)$
with respect to $\epsilon$ evaluated at $\epsilon=0.$
The bifurcation function for  the system $F_\epsilon$
adapted to the period manifold ${\cal A}$ is the function
${\cal B}:{\bfsp R}^k\to {\bfsp R}^k$
defined by
\[
{\cal B}(\theta)=H(\theta)
\left(
\begin{array}{c}
0 \\
{\cal N}(\theta)\\
{\cal M}(\theta)
\end{array}
\right)
\]
where
\begin{eqnarray*}
{\cal M}(\theta)&:=&\int_0^{T(\theta)}
b^{-1}(s,\theta)G^{\mbox{nor}}(s,\theta)\,ds,  \\
{\cal N}(\theta)&:=&\int_0^{T(\theta)}
c^{-1}(s,\theta)G^{\mbox{tan}}(s,\theta)-
c^{-1}(s,\theta)a(s,\theta)b^{-1}(s,\theta)
G^{\mbox{nor}}(s,\theta)\,ds
\end{eqnarray*}
and where $T(\theta)$ denotes the time of first return to the Poincar\'e
section
for the unperturbed solution $t\mapsto x(t,\theta).$
The following theorem is proved in \cite{cccc}.
\begin{theorem}
Suppose $F_\epsilon$ given by
\[
\dot x=f(x)+\epsilon g(x,\dot x,\epsilon),
\;\;x\in{\bfsp R}^{n+1},\;\;\epsilon\in {\bfsp R}
\]
has a normally nondegenerate period manifold ${\cal A}$ with adapted coordinate
system given by $(s,\theta,\zeta)\mapsto \Delta(s,\theta,\zeta).$
If $\theta_0$ is a simple zero of the bifurcation function
$\theta\to {\cal B}(\theta)$
adapted to ${\cal A},$ then there is an $\epsilon_*>0$ and a smooth function
$\beta:(-\epsilon_*,\epsilon_*)\to {\bfsp R}^k\times {\bfsp R}^{n-k}$
with $\beta(0)=(\theta_0,0)$ such that $\Delta(0,\beta(\epsilon))$ is the
initial value for a periodic solution of $F_\epsilon.$
\end{theorem}
\section{Persistent periodic solutions of the coupled oscillator}
In this section we apply the results outlined in \S 2 to the system
$E_\epsilon$
defined in the introduction. To do this we must identify the bifurcation
function. Other, perhaps simpler examples of the identification procedure are
given
in \cite{cccc}.
In any case, there are several steps.

Step 1. [Definition of the period manifold]
Under the assumptions $1$--$3$ listed in the introduction, the
unperturbed system $E_0$ has a three dimensional period manifold given by
${\cal A}:=A\times \Gamma.$ In fact, every solution of the unperturbed system
starting on ${\cal A}$ has the same period $T_{\cal A}:=K_1\eta_1.$

Step 2. [Adapted coordinates]
For  vectors  $v=(v_1,v_2)$ and $w=(w_1,w_2)$ in ${\bfsp R}^2,$  let
$<v,w>$ denote the usual inner product, $||v||^2:=<v,v>,\,$
$v^\perp:=(-v_2,v_1)$ and $v\wedge w:=<w,v^\perp>.$ Using these definitions
and the unperturbed vector fields $f_1$ and $f_2$ on ${\bfsp R}^2,$
we define two smooth
vector fields $f_1^\perp$ and $f_2^\perp$ on ${\bfsp R}^2.$
Also, we let $\varphi^i$ denote
the flow of $\dot x_i=f_i(x_i)$ and $\psi^i$ denote the flow of
$\dot x_i=f_i^\perp(x_i)$ for $i=1,2.$
For each $x=(x_1,x_2)$ in $A\times \Gamma,$ we define a splitting over ${\cal
A}$
by
\begin{eqnarray*}
{\cal E}(x)&=&
\left [
\left (
\begin{array}{c}
f_1(x_1)  \\
f_2(x_2)
\end{array}
\right )
\right ], \\
{\cal E}^{\mbox{tan}}(x)&=&
\left [
\begin{array}{lr}
\left (
\begin{array}{c}
||f_1(x_1) ||^{-2}f_1^\perp(x_1)  \\
0
\end{array}
\right ), &
\left (
\begin{array}{c}
0\\
f_2(x_2)
\end{array}
\right )
\end{array}
\right],   \\
{\cal E}^{\mbox{nor}}(x)&=&
\left [
\left (
\begin{array}{c}
0\\
||f_2(x_2)||^{-2}f_2^\perp(x_2)
\end{array}
\right )
\right ]
\end{eqnarray*}
where the square brackets here and hereafter denote the
subspace spanned by the enclosed vectors.
This gives
\[
T_x{\cal A}={\cal E}(x)\oplus{\cal E}^{\mbox{tan}}(x)\oplus
{\cal E}^{\mbox{nor}}(x).
\]
Next, fix $\xi_1\in A$ and $\xi_2\in \Gamma$ and define adapted coordinates
$
\Delta:{\bfsp R}^4 \to {\bfsp R}^4
$
by
\[
(s,p,q,\zeta)\mapsto \left(\varphi^1_s(\psi^1_p(\xi_1)),
\psi^2_\zeta(\varphi^2_{s+q}(\xi_2))\right).
\]
If
\[
\Sigma_0:=\left\{
\Delta(0,p,q,\zeta)\;\left |\right.\; (p,q,\zeta)\in {\bfsp R}^3
\right\},
\]
then there is some open subset $\Sigma\subseteq \Sigma_0$  that
is a three dimensional Poincar\'e section for $E_0$ at $(\xi_1,\xi_2).$

Step 3. [Fundamental matrix of variational equation in adapted coordinates]
We consider the fundamental matrix solution $\Phi(t)$ with initial condition
$\Phi(0)=I$ for the variational equation
\[
\left(
\begin{array}{c}
\dot w_1 \\
\dot w_2
\end{array}
\right)
=\left(
\begin{array}{cc}
Df_1\left(\varphi^1_t(\psi^1_p(\xi_1))\right) & 0 \\
0 & Df_2\left(\varphi^2_{t+q}(\xi_2)\right)
\end{array}
\right)
\left(
\begin{array}{c}
 w_1 \\
 w_2
\end{array}
\right)
\]
and recall  Diliberto's theorem \cite{cccfe,cccc,dili}.
\begin{theorem}[Diliberto's Theorem \cite {cccfe,dili}]
If $\dot x=f(x),\;$ $x\in {\bfsp R}^2,\;$ $f(\xi)\ne 0,$ and
$t\mapsto x(t,p)$ is the solution of the differential equation such that
$x(0,p)=p,$ then the homogeneous variational equation
\[\dot W=Df(x(t,\xi))W\]
has a fundamental matrix solution $t\mapsto\Psi(t)$
\[
\Psi(t)=
\left(
\begin{array}{cc}
1&\alpha(t,\xi)  \\
0&\beta(t,\xi)
\end{array}
\right)
\]
with respect to the  moving frame
\[
\left \{
f(t,\xi),\; ||f(t,\xi)||^{-2}{f^\perp}(t,\xi)
\right \}
\]
where
\begin{eqnarray*}
f(t,\xi)&:=& f(x(t,\xi)) \\
\beta(t,\xi)&=&\exp \int_0^t \div f(s,\xi)\,ds,   \\
\alpha(t,\xi)&=&\int_0^t
\left \{ \frac{1}{||f||^2}(2\kappa ||f||-\curl f)\beta\right\}(s,\xi)\,ds
\end{eqnarray*}
and $\kappa$ denotes the signed scalar curvature
\[
\kappa(t,\xi):=\frac{1}{||f(t,\xi)||^3}f(t,\xi)\wedge Df(t,\xi)f(t,\xi).
\]
\end{theorem}
Also, to compress the notation,
we define
\[
\begin{array}{ll}
\alpha_1(s,p):=\alpha_1(s,\psi^1_p(\xi_1)), &
\beta_1(s,p):=\beta_1(s,\psi^1_p(\xi_1)), \\
f_1(s,p):=f_1(\varphi^1_s(\psi^1_p(\xi_1))),&
f_1^\perp(s,p):=f_1^\perp(\varphi^1_s(\psi^1_p(\xi_1))), \\
\alpha_2(s,q):=\alpha_2(s,\varphi^2_q(\xi_2)), &
\beta_2(s,q):=\beta_2(s,\varphi^2_q(\xi_2)), \\
f_2(s,q):=f_2(\varphi^2_{s+q}(\xi_2)),&
f_2^\perp(s,p):=f_2^\perp(\varphi^2_{s+q}(\xi_2))
\end{array}
\]
where the subscripts on $\alpha$ and $\beta$  refer to the functions as defined
in Diliberto's theorem for the unperturbed equations $\dot x_i=f_i(x_1),\,$
$i=1,2.$
Now, the fundamental matrix solution relative to the basis  ${\cal S}$ for our
splitting
\begin{eqnarray*}
\lefteqn{\left\{
F(t,p,q),F_1^{\mbox{tan}}(t,p),F_2^{\mbox{tan}}(t,q),F^{\mbox{nor}}(t,q)
\right \}:=}&&  \\
&&\left\{
\left(
\begin{array}{c}
f_1(t,p) \\
f_2(t,q)
\end{array}
\right),
\left(
\begin{array}{c}
||f_1(t,p)||^{-2}f_1^\perp(t,p) \\
0
\end{array}
\right),
\left(
\begin{array}{c}
0 \\
f_2(t,q)
\end{array}
\right),
\left(
\begin{array}{c}
0 \\
||f_2(t,q)||^{-2}f_2^\perp(t,q)
\end{array}
\right)
\right\}
\end{eqnarray*}
is given by
\[
\Phi(t)=
\left (
\begin{array}{cccc}
1 & \alpha_1(t,p) & 0 & 0 \\
0 & \beta_1(t,p) & 0 & 0  \\
0 & 0 & 1 & \alpha_2(t,q) \\
0 & 0 & 0 & \beta_2(t,q)
\end{array}
\right).
\]
This means the associated maps $a,$ $b$ and $c$
defined in \S 2 reduce as follows:
\newline
$
a: {\cal E}^{\mbox{nor}}(\psi^1_p(\xi_1),\varphi^2_q(\xi_2))\to
{\cal
E}^{\mbox{tan}}(\varphi^1_t(\psi^1_p(\xi_1)),\varphi^1_t(\varphi^2_q(\xi_2)))
$
is given by the  $2\times 1$ matrix
\[
\left(
\begin{array}{c}
0 \\
\alpha_2(t,q)
\end{array}
\right),
\]
$
b:{\cal E}^{\mbox{nor}}(\psi^1_p(\xi_1),\varphi^2_q(\xi_2))\to
{\cal
E}^{\mbox{nor}}(\varphi^1_t(\psi^1_p(\xi_1)),\varphi^1_t(\varphi^2_q(\xi_2)))
$
is given by the $1\times 1$ matrix
$(\beta_2(t,q))$ and
$
c:{\cal E}^{\mbox{tan}}(\psi^1_p(\xi_1),\varphi^2_q(\xi_2))\to
{\cal
E}^{\mbox{tan}}(\varphi^1_t(\psi^1_p(\xi_1)),\varphi^1_t(\varphi^2_q(\xi_2)))
$
is given by the $2\times 2$ matrix
\[
\left(
\begin{array}{cc}
\beta_1(t,p)    &    0 \\
     0          &    1
\end{array}
\right).
\]

Step 4 [Normal nondegeneracy] Define the transit time map $T:{\bfsp R}^3\to
{\bfsp R}$
given by $(p,q,\zeta)\mapsto T(p,q,\zeta)$ where $T(p,q,\zeta)$ denotes the
time of
first return of the point $\Delta(0,p,q,\zeta)\in \Sigma$ to $\Sigma_0$ and
note
$T(p,q,0)\equiv T_{\cal A}.$ To show the normal nondegeneracy
we must show the kernel of the derivative of the displacement at each point
on $\xi\in \Sigma \cap {\cal A}$  is two dimensional. In the
present case, since we already know the kernel contains the subspace
${\cal E}^{\mbox{tan}}(\xi),$  it suffices to show
the derivative of the Poincar\'e map at $\xi$ is not the identity.
To prove this we show
\[
DP\left(\psi^1_p(\xi_1),\varphi^2_q(\xi_2),0\right)
\left(
\begin{array}{c}
0   \\
f_2^\perp(0,q)
\end{array}
\right)
\ne
\left(
\begin{array}{c}
0   \\
f_2^\perp(0,q)
\end{array}
\right).
\]
The vector in the last formula is tangent to the curve
\[\zeta\mapsto \left(\psi^1_p(\xi_1),\psi^2_\zeta(\varphi^2_q(\xi_2)\right)\]
at $\zeta=0.$ So, we must compute the tangent to the curve
\[
\zeta\mapsto P\left(\psi^1_p(\xi_1),\psi^2_\zeta(\varphi^2_q(\xi_2))\right)
=
\left(
\varphi^1_{T(p,q,0)}(\psi^1_p(\xi_1)),
\varphi^2_{T(p,q,0)}(\psi^2_\zeta(\varphi^2_q(\xi_2)))
\right)
\]
at $\zeta=0.$ The computation is just an application of Diliberto's theorem.
In fact, we obtain
\begin{eqnarray*}
\lefteqn{DP\left(\psi^1_p(\xi_1),\varphi^2_q(\xi_2),0\right)
\left(
\begin{array}{c}
0   \\
f_2^\perp(0,q)
\end{array}
\right)=
\left(
\begin{array}{c}
0  \\
D\varphi^2_{T(p,q,0)}(\varphi^2_q(\xi_2)f_2^\perp(0,q)
\end{array}
\right)}&& \\
&=&
\left(
\begin{array}{c}
0  \\
||f_2(0,q)||^2
\left(
\alpha_2(T(p,q,0),q)f_2(0,q)+\beta_2(T(p,q,0),q)||f_2(0,q)||^2f_2^\perp(0,q)
\right)
\end{array}
\right).
\end{eqnarray*}
The infinitesimal displacement of our vector is given by
\begin{eqnarray*}
\lefteqn{{\cal R}(q):=DP(\psi^1_p(\xi_1),\varphi^2_q(\xi_2),0)
\left(
\begin{array}{c}
0   \\
f_2^\perp(0,0)
\end{array}
\right)
-
\left(
\begin{array}{c}
0   \\
f_2^\perp(0,0)
\end{array}
\right)} \mbox{\hspace*{.5in}}&&                                  \\
&=&
\left(
\begin{array}{c}
0  \\
||f_2(0,q)||^2
\alpha_2(T_{\cal A},q)f_2(0,q)+(\beta_2(T_{\cal A},q)-1)f_2^\perp(0,q)
\end{array}
\right).
\end{eqnarray*}
To see that ${\cal R}(q)\ne 0,$  we use the following facts:
$\beta_2(T_{\cal A},q)$ is the characteristic multiplier of $\Gamma$
and the derivative of the transit
time function at $\Gamma$ is  given by $-||f_2(0,0)||\alpha_2(T_{\cal A},q),$
see \cite{cccfe} or \cite{cccc} for more explanation. Since, by the hypotheses
stated in \S 1, either $\Gamma$ is hyperbolic or $\Gamma$ belongs to a period
annulus such that the derivative of a period function does not vanish at
$\Gamma,$
it follows that ${\cal A}$ is normally nondegenerate.

Step 5. [Projection to Complement of the range of $D\delta(p,q,0,0)$]
It is clear from step $4$ that a two dimensional complement for the
range of $D\delta(p,q,0,0),$ expressed with respect to the basis ${\cal S}$
for the splitting over
${\cal A},$ is given by
\[\left\{F_1^{\mbox{tan}}(0,p),{\cal R}^\perp(q)\right\}\]
where
\begin{eqnarray*}
{\cal R}^\perp(q):=
\left(
\begin{array}{c}
0   \\
(1-\beta_2(T(p,q,0),q)f_2(0,q)+
||f_2(0,q)||^2 \alpha_2(T_{\cal A},q)f_2^\perp(0,q)
\end{array}
\right).
\end{eqnarray*}
Moreover, since
\[\left\{
F(0,p,q), F_1^{\mbox{tan}}(0,p),
{\cal R}(q),{\cal R}^\perp(q)
\right\}
\]
is a basis ${\cal T}$ for ${\bfsp R}^4,$
the projection from the original splitting to the chosen complement
for the range is easy to compute.
In fact, there are four functions, each mapping ${\bfsp R}$ to ${\bfsp R},$
given by
$q\mapsto k_1(q),$ $q\mapsto k_2(q),$ $q\mapsto B(q)$
and $q\mapsto C(q)$ such that
\begin{eqnarray*}
F_2^{\mbox{tan}}(0,q)&=& k_1(q){\cal R}(q)+B(q){\cal R}^\perp(q), \\
F^{\mbox{nor}}(0,q)&=& k_2(q){\cal R}(q)+C(q){\cal R}^\perp(q). \\
\end{eqnarray*}
Thus, the matrix of the required projection
\[
H(p,q):\left({\cal E}\oplus{\cal E}^{\mbox{tan}}\oplus {\cal
E}^{\mbox{nor}}\right)
(\Delta(0,p,q,0))\to {\bfsp R}^2
\]
with respect to the (ordered) basis ${\cal S}$ on its domain and the
(ordered) basis ${\cal T}$ on its range
is given by the linear map
\[
H(p,q)
\left(
\begin{array}{c}
\varepsilon \\
\tau_1\\
\tau_2\\
\eta
\end{array}
\right)
=
\left(
\begin{array}{c}
\tau_1\\
B(q)\tau_2+C(q)\eta
\end{array}
\right)
\]
where
\begin{eqnarray*}
B(q)&:=&\frac{1-\beta_2(T_{\cal A},q)}
{||f_2(0,q)||^4\alpha_2(T_{\cal A},\xi_2)^2+
\left(1-\beta_2(T_{\cal A},q)\right)^2}, \\
C(q)&:=&\frac{\alpha_2(T_{\cal A},q)}
{||f_2(0,q)||^4\alpha_2(T_{\cal A},\xi_2)^2+
\left(1-\beta_2(T_{\cal A},q)\right)^2}.
\end{eqnarray*}

Step 6. [Adapted Components for perturbation]
The derivative with respect to $\epsilon$ at $\epsilon=0$ of the vector field
associated with $E_\epsilon$ along the unperturbed solution is given by
\[
G(t,p,q):=
\left(
\begin{array}{c}
g_1(t,p,q) \\
g_2(t,p,q)
\end{array}
\right)
:=
\left(
\begin{array}{c}
g_1(x_1(t,p),\dot x_1(t,p),x_2(t,q),\dot x_2(t,q),0) \\
g_2(x_1(t,p),\dot x_1(t,p),x_2(t,q),\dot x_2(t,q),0)
\end{array}
\right)
\]
where $t\mapsto (x_1(t,p),x_2(t,q))$
is the unperturbed solution starting at $\Delta(0,p,q,0).$
The vector $G(t,p,q)$ has a unique expression as
a linear combination of the vectors in the basis ${\cal S}.$
In fact, we suppose
\[
G(t,p,q)=\varepsilon F(t,p,q)+\tau_1 F_1^{\mbox{tan}}(t,p)
+\tau_2 F_2^{\mbox{tan}}(t,q)+\eta F^{\mbox{nor}}(t,q)
\]
and compute inner products
with respect to $f_1,$ $f_1^\perp,$
$f_2$ and $f_2^\perp$ to obtain
\begin{eqnarray*}
G^{\mbox{tan}}(t,p,q)&:=&
\left(
\begin{array}{c}
\tau_1(t,p,q) \\
\tau_2(t,p,q)
\end{array}
\right),                \\
G^{\mbox{nor}}(t,p,q)&:=&\eta(t,p,q),
\end{eqnarray*}
where
\begin{eqnarray*}
\tau_1(t,p,q)&=&f_1(t,p)\wedge g_1(t,p,q),\\
\tau_2(t,p,q)&=&\frac{1}{||f_2(t,q)||^2}\left<g_2(t,p,q),f_2(t,q)\right>-
\frac{1}{||f_1(t,p)||^2}\left<g_1(t,p,q),f_1(t,p)\right>, \\
\eta(t,p,q)&=&f_2(t,q)\wedge g_2(t,p,q).
\end{eqnarray*}

Step 7. [Bifurcation function]
Using the definitions of \S 2 and the results of steps $3$--$4$ we now have
\[
{\cal M}(p,q)=\int_0^{T_{\cal A}}b^{-1}(s,q)G^{\mbox{nor}}(s,p,q)\, ds
\]
given by
\[\int_0^{T_{\cal A}}\frac{1}{\beta_2(s,q)}f_2(t,q)\wedge g_2(t,p,q)\, ds, \]
and
\begin{eqnarray*}
{\cal N}(p,q)&=&\int_0^{T_{\cal A}}c^{-1}(t,p)G^{\mbox{tan}}(s,p,q)
-c^{-1}(t,p)a(t,q)b^{-1}(t,q) G^{\mbox{nor}}(s,p,q)\, ds \\
&:=&
\left(
\begin{array}{c}
{\cal N}_1(p,q) \\
{\cal N}_2(p,q)
\end{array}
\right)
\end{eqnarray*}
given by
\begin{eqnarray*}
{\cal N}_1(p,q)&=&
\int_0^{T_{\cal A}}\frac{1}{\beta_1(s,p)}f_1(t,p)\wedge g_1(t,p,q)\, ds, \\
{\cal N}_2(p,q)&=&\int_0^{T_{\cal A}}
\frac{1}{||f_2(t,q)||^2}\left<g_2(t,p,q),f_2(t,q)\right>\\
&&-\frac{1}{||f_1(t,p)||^2}\left<g_1(t,p,q),f_1(t,p)\right>
-\frac{\alpha_2(t,q)}{\beta_2(t,q)}f_2(t,q)\wedge g_2(t,p,q)\, ds.
\end{eqnarray*}
Thus, the bifurcation function is given by
\[
{\cal B}(p,q)=H(p,q)
\left(
\begin{array}{c}
0 \\
{\cal N}_1(p,q) \\
{\cal N}_2(p,q) \\
{\cal M}(p,q)
\end{array}
\right)
=
\left(
\begin{array}{c}
{\cal N}_1(p,q)\\
B(q){\cal N}_2 (p,q)+C(q){\cal M}(p,q)
\end{array}
\right).
\]
In practice, it is more convenient to clear the nonzero denominator of the
second component and to use the normalized bifurcation function given by
\[
{\cal C}(p,q):=
\left(
\begin{array}{c}
{\cal N}_1(p,q)\\
\left(1-\beta_2(T_{\cal A},q)\right){\cal N}_2 (p,q)
+\alpha_2(T_{\cal A},q){\cal M}(p,q)
\end{array}
\right).
\]
Of course, ${\cal C}$ and ${\cal B}$ have the same set of simple zeros.
\section{Applications}
We consider an application to the model equations for
a flywheel attached to an elastic support as described in \cite {sv}.
The model equations are typical for resonance phenomena
and are given by
\begin{eqnarray*}
\ddot z+\omega^2 z&=&\frac{\epsilon}{m}
\left(-f(z)-\beta \dot z+q_1\dot \theta^2 \cos{\theta}\right)+O(\epsilon^2), \\
\ddot \theta&=& \epsilon
\left( \frac{1}{J_0}M_1(\dot \theta)+q_2\,\mbox{g}\sin{\theta}-
q_2\omega^2 z\sin{\theta}\right) +O(\epsilon^2),
\end{eqnarray*}
where
$x$ denotes the displacement of the flywheel relative to its support,
$\theta$ denotes the angular position of the rotating flywheel relative
to the (upward) vertical, $\mbox{g}$ is the gravitational constant and
$M_1$ is the motor characteristic. The remaining
parameters are all constant with, of course, $\epsilon$ being a small
parameter.
To apply the results of \S 3, we write the model equation as a first order
system using the transformation $x=\dot\theta\cos{\theta},$
$y=\dot\theta\sin{\theta},$ and assuming $\dot \theta >0$ to obtain
\begin{eqnarray*}
\dot z&=&-\omega w,  \\
\dot w&=& \omega z-\epsilon g(z,w,x,y), \\
\dot x&=& -y\sqrt{x^2+y^2}+\epsilon\frac{x}{\sqrt{x^2+y^2}}h(z,w,x,y)
       +O(\epsilon^2), \\
\dot y&=& x\sqrt{x^2+y^2}+\epsilon\frac{y}{\sqrt{x^2+y^2}}h(z,w,x,y)
       +O(\epsilon^2),
\end{eqnarray*}
where
\begin{eqnarray*}
g(z,w,x,y)&=& \frac{1}{m\omega}\left(-f(z)-\beta w+q_1x\sqrt{x^2+y^2}\right),
\\
h(z,w,x,y)&=&  \frac{1}{J_0}M_1\left(\sqrt{x^2+y^2}\right)
   +q_2\,\mbox{g}\frac{y}{\sqrt{x^2+y^2}}-q_2\omega^2
\frac{yz}{\sqrt{x^2+y^2}}.
\end{eqnarray*}
The transformation to $(x,y)$ variables is geometrically a coordinate
chart on the tangent bundle of the circle
\[
\{(\theta,\dot \theta)\left |\right.
\theta\in{\bfsp S}^1\;\mbox{and}\;\dot\theta\in {\bfsp R}\}.
\]
The chart does not contain the
zero section ($\dot\theta=0$), but this set is not near the resonance.
In fact, the first oscillator is linear with its period annulus $A$ having
period $2\pi/\omega$ while the second oscillator has a period annulus
at the origin whose period function is given by $r\mapsto 2\pi/r$ where
$r:=\sqrt{x^2+y^2}.$ The primary resonance is given by $r=\omega.$ In
other words, the resonant periodic solution $\Gamma$
in the second oscillator lies on the invariant circle of radius $\omega.$
With
\[
f_1(z,w):=-w\frac{\partial}{\partial z}+z\frac{\partial}{\partial
w},\hspace{.3in}
f_2(x,y):=-y\sqrt{x^2+y^2}\frac{\partial}{\partial x}+
x\sqrt{x^2+y^2}\frac{\partial}{\partial y},
\]
$\xi_1:=(1,0)$ and  $\xi_2:=(\omega,0),$ we find
the solution of the unperturbed system with initial value
$(\phi^1_p(\xi_1),\varphi^2_q(\xi_2))$ to be
given by
\[
\begin{array}{lclccl}
z(t,p)&=&\mbox{e}^{-p}\cos(\omega t),
\hspace*{.5in} &w(t,p)&=&\mbox{e}^{-p}\sin(\omega t),  \\
x(t,q)&=& \omega\cos(\omega(t+q)),\hspace*{.5in}
&y(t,q)&=&\omega\sin(\omega(t+q)).
\end{array}
\]
{}From the results of \S 3,  the bifurcation function is
\begin{eqnarray*}
{\cal B}(p,q)&=&\left(
\omega \int_0^{2\pi/\omega}wg(z,w,x,y)\,dt,\,
-\int_0^{2\pi/\omega}(x^2+y^2)h(z,w,x,y)\,dt
\right) \\
&=& \left(
\frac{\pi\omega}{m}\,\mbox{e}^{-2p}(\beta-q_1\omega\,\mbox{e}^{p}\sin(\omega
q)),\,
-\frac{\pi\omega^2}{J_0}(2 M_1(1/\omega)
   -J_0q_2\omega^2 \,\mbox{e}^{-p}\sin(\omega q))\right).
\end{eqnarray*}
The bifurcation function has either $0,$ $1$ or $2$ zeros depending on the
values of the parameters. The zeros are obtained when the following two
equations can be solved for both $p$ and $q:$
\[
\mbox{e}^{2p}=\frac{\beta\omega J_0 q_2}{2 M_1(1/\omega) q_1},\hspace{.3in}
\sin{\omega q}=\frac{\beta}{\omega q_1}\sqrt{\frac{2 M_1(1/\omega)q_1}
{\beta\omega J_0q_2}}.
\]
If we choose $\beta=\omega=J_0=q_1=q_2=1$ and $M_1(1)=1/4,$ then
these equations reduce to
\[\mbox{e}^{p}=\sqrt{2},\hspace{.5in} \sin{q}=\frac{1}{\sqrt{2}}. \]
In this case $p=\ln \sqrt{2},\;$ $q={\pi}/{4},\, {3\pi}/{4}$
are zeros of the bifurcation function. Since the bifurcation function
can be normalized to
\[
(p,q)\mapsto
\left(
\beta-q_1\omega\,\mbox{e}^{p}\sin(\omega q),\,
2 M_1(1/\omega) -J_0q_2\omega^2 \,\mbox{e}^{-p}\sin(\omega q)
\right),
\]
a map whose Jacobian is
\[2q_1q_2J_0\omega^2\sin(\omega q)\cos(\omega q),\]
its zeros are simple except when $\sin(\omega q)=\pm 1.$
In particular, the zeros of the numerical example are simple and,
by the results of \S 3,
there are two bifurcating families of periodic solutions in the
model equations for the flywheel with elastic support. We emphasize
that although the analysis uses only the $O(\epsilon)$ terms of the
model, our result is valid for small $\epsilon$ for the full model equations,
compare \cite{sv}.

The analysis just given is prototypical. However, there are other resonances
to consider. Using the notation defined above, the general resonance relation
is given by
\[ K_1\frac{2\pi}{\omega}=K_2\frac{2\pi}{r}\]
or $r=K_2 \omega/K_1.$
On the resonant orbit
\[
x(t,q)=\frac{K_2}{K_1}\omega\cos{\frac{K_2}{K_1}\omega(t+q)},\hspace{.4in}
y(t,q)=\frac{K_2}{K_1}\omega\sin{\frac{K_2}{K_1}\omega(t+q)}.
\]
Thus, the first component of the bifurcation function is
given by
\begin{eqnarray*}
\lefteqn{\omega\int_0^{K_12\pi/\omega}wg(z,w,x,y)\,dt}&&  \\
&=&\frac{1}{m}\int_0^{K_1 2\pi/\omega}\,\mbox{e}^{-p}\sin{\omega t}
\left(
-f(z)-\beta \,\mbox{e}^{-p}\sin{\omega
t}+q_1\left(\frac{K_2}{K_1}\omega\right)^2
 \cos{\frac{K_2}{K_1}\omega (t+q)}
\right)\, dt  \\
&=& \frac{1}{m}\,\mbox{e}^{-2p}\frac{K_1\pi}{\omega}
+\frac{q_1}{m}\left(\frac{K_2}{K_1}\omega\right)^2\,\mbox{e}^{-p}I_0(q)
\end{eqnarray*}
where
\[I_0(q):=\int_0^{K_1 2\pi/\omega}\sin{\omega t}\cos{\frac{K_2}{K_1}\omega
(t+q)}\,dt.\]
As $I_0(q)$ is nonzero only when $K_1=K_2,$
nondegenerate bifurcation to periodic orbits occurs only for the primary
resonance.

Up to this point we have assumed
several forces are small. To illustrate the possibility of relaxing this
hypothesis consider the rotor to be influenced strongly by a ``gravitational''
force. It is convenient to measure the inclination of the rotor
by the angle of displacement from the direction of the gravitational force,
downward
vertical, i.e., we use the angle $\psi=-\theta-\pi.$
The model equations (up to first order in $\epsilon$) become
(to first order)
\begin{eqnarray*}
\ddot z+\omega^2 z&=&\frac{\epsilon}{m}
\left(-f(z)-\beta \dot z+q_1\dot \psi^2 (-\cos{\psi})\right), \\
\ddot \psi&=& -\epsilon
\left( \frac{1}{J_0}M_1(-\dot \psi)+q_2\,{\rm g}\sin{\psi}-
q_2\omega^2 z\sin{\psi}\right).
\end{eqnarray*}
To study the  strong gravitational
effect we assume ${\rm g}:=G/\epsilon$ and transform
the independent variable by $\tau=t\sqrt{q_2 G}$ to obtain
\begin{eqnarray*}
q_2 G z''+\omega^2 z&=& -\frac{\epsilon}{m}\left(
 f(z)+\beta z' \sqrt{q_2 G}  +q_1q_2G(\psi')^2\cos{\psi}\right), \\
q_2 G \psi '' +q_2 G \sin {\psi}& =&-\epsilon\left (
\frac{1}{J_0} M_1(-\psi' \sqrt{q_2 G})-q_2\omega ^2 z\sin{\psi}\right),
\end{eqnarray*}
which we rewrite in the form
\begin{eqnarray*}
\ddot z+\Omega^2 z &=&\epsilon g(z,\dot z,\theta,\dot \theta), \\
\ddot \theta+\sin{\theta} &=&\epsilon h(z,\dot z,\theta,\dot \theta), \\
\end{eqnarray*}
where
\begin{eqnarray*}
g(z,\dot z,\theta,\dot \theta)&:=& -\left (
F(z)+\lambda \dot z +A\dot \theta^2\cos{\theta}
\right),   \\
h(z,\dot z,\theta,\dot \theta)&:=& -\left(
M(\dot \theta)-B z \sin{\theta}
\right),
\end{eqnarray*}
and the new parameters and functions have the obvious meaning.
In particular,
\[
\lambda:=\frac{\beta}{m}(q_2G)^{-1/2},\hspace{.2in}
\Omega:=\omega (q_2G)^{-1/2},\hspace{.2in}
A:=\frac{q_1}{m},\hspace{.2in} B:=\frac{\omega^2}{G}.
\]
The first oscillator, corresponding to the elastic support, has the entire
punctured phase plane as an isochronous period annulus with period
$2\pi/\Omega.$ In
fact, if we view the system in the phase plane as
\[
\dot z=-\Omega w,\hspace{.5in}
\dot w=\Omega z-\frac{\epsilon}{\Omega}g(z,\dot z,\theta,\dot \theta),
\]
the solution of the unperturbed oscillator with initial value
$(\mbox{e}^{-p},0)$ is given by
\[
z(t,p)=\mbox{e}^{-p}\cos{\Omega t},\hspace{.5in}
w(t,p)=\mbox{e}^{-p}\sin{\Omega t}.
\]
The second oscillator models  the rotor influenced by a gravitational field.
The
unperturbed second oscillator is a mathematical pendulum.
It has a period annulus (with strictly
monotone period function) surrounding the origin of the phase plane. This
period
annulus corresponds to the nonrotational oscillations of the pendulum. Also,
there is a period annulus in the phase cylinder (with strictly monotone
period function) corresponding to the rotational oscillations. Thus we have
the hypotheses required to apply the theoretical results  of \S 3.
The analysis to follow  uses elliptic functions. Perhaps
this can be avoided?

To compute the bifurcation function we require the time dependent solutions of
the
mathematical pendulum given in the phase plane  by the first order system
\[\dot \theta=v,\hspace{.5in} \dot v=-\sin{\theta}.\]
For the convenience of the reader and to fix notation we outline the usual
derivation.
We begin by considering the period annulus in the phase plane.
The mathematical pendulum has the first integral $I:=v^2/2-\cos{\theta}.$
For a periodic trajectory $\Gamma$ let $(a,0)$ denote the coordinates of its
intersection with the $\theta$-axis. On $\Gamma$ the energy is $I\equiv -\cos
a$
and $\dot \theta^2=2(\cos{\theta}-\cos {a}).$ By integration and the change of
variables $\sin({\theta/2})=\sin({a/2})\sin{\varphi}$ we find
\[t=\int_0^{\varphi(t)}\frac{1}{\sqrt{1-\sin^2{(a/2)}\sin^2{s}}}\, ds \]
where $\theta(t)$ is the solution of the mathematical pendulum with the
initial value
\[(\theta(0),\dot \theta(0))=(0,2\sin{(a/2)}).\]
Or, in terms of Jacobian elliptic functions, cf. \cite{bf,ww}, where
the elliptic modulus is $k:=\sin{(a/2)},$
we find
\[
\sin{\varphi(t)}=\mbox{sn}[t,k]\]
and, using the trigonometric double angle formulas,
\[\cos{\theta(t)}=1-2k^2\,\mbox{sn}^2[t,k].\]
Also, the period of $\Gamma$ is given by
\[
4\int_0^{\pi/2}\frac{1}{\sqrt{1-k^2\sin^2{s}}}\,ds=4K(k)
\]
where $K(k)$ is the complete elliptic integral of the first kind. Since,
$t\mapsto \mbox{sn}(t)$ has real period $4K $ (here and hereafter if the
elliptic modulus is not given explicitly it is understood to be
$k=\sin{(a/2)}),$  the periodic orbit $\Gamma$ is resonant
when there are relatively prime
positive integers $K_1 $ and $K_2$ such that
\[K_1\frac{2\pi}{\Omega}=K_2 4 K(k).\]
Under this assumption and in view of the first order system
\begin{eqnarray*}
\dot z&=&-\Omega w,\\
\dot w&=&\Omega z-\frac{\epsilon}{\Omega}g,\\
\dot \theta&=& v, \\
\dot v&=& -\sin{\theta} +\epsilon h,
\end{eqnarray*}
the bifurcation function for a nonrotational resonance is given by
\[
{\cal B}(p,q)=\left(
\int_0^{K_12\pi/\Omega}wg\,dt,\hspace{.1in} \int_0^{K_12\pi/\Omega}vh\,dt
\right)
\]
where $q$ is the coordinate on $\Gamma$ introduced by using the
solution $t\mapsto \theta(t+q),$ for $0\le q <4K(k).$ The components of the
bifurcation function are computed as follows:
\begin{eqnarray*}
\int_0^{K_12\pi/\Omega}wg\,dt&=& K_1\pi\lambda
\,\mbox{e}^{-2p}-A\,\mbox{e}^{-p}I_1(q), \\
\int_0^{K_12\pi/\Omega}vh\,dt&=& B\,\mbox{e}^{-p}I_2(q)-I_3(q),
\end{eqnarray*}
where
\begin{eqnarray*}
I_1(q)&:=&
\int_0^{K_12\pi/\Omega}(\dot\theta(t+q))^2\cos{\theta(t+q)}\sin{\Omega
t}\,dt,\\
I_2(q)&:=&
\int_0^{K_12\pi/\Omega}\dot\theta(t+q)\sin{\theta(t+q)}\cos{\Omega t}\,dt, \\
I_3(q)&:=&
\int_0^{K_12\pi/\Omega}\dot\theta(t+q)M(\dot\theta(t+q))\,dt.
\end{eqnarray*}

The integral $I_3$ depends on the static characteristic of the motor
and the damping associated with the rotational motion as encoded in
the function $M.$ As a typical example and for definiteness in
the computation we take $M$ to be linear,
\[M(\dot \theta):=m_1+m_2\dot \theta;\]
more general model functions can be handled in
a similar manner. For the linear case, we have the following proposition.
\begin{eqnarray*}
I_3(q)&=&\int_0^{K_12\pi/\Omega}m_1\dot \theta+m_2\dot \theta^2\,dt\\
&=& 2m_2\int_0^{K_12\pi/\Omega}\cos{\theta(t+q)}-\cos{a}\,dt \\
&=& -4m_2K_1\frac{\pi}{\Omega}\cos {a}+2m_2\int_0^{K_24K}\cos{\theta(t)}\,dt\\
&=& -2m_2K_24K\cos{a}+2m_2\int_0^{K_24K}1-2k^2\,\mbox{sn}^2(t)\,dt.
\end{eqnarray*}
The formula
{\bf 310.02} of \cite {bf} can be used to evaluate the integral with integrand
$\mbox{sn}^2(t)$ to obtain
\begin{eqnarray*}
I_3(q)&=&-8m_2K_2(1+\cos{a})K(k) +4m_2E(\mbox{am}\,[K_24K(k),k],k)\\
&=&-8m_2K_2(1-k^2)K(k) +4m_2E(\mbox{am}\,[K_24K(k),k],k) \\
\end{eqnarray*}
where $E(\varphi,k)$ denotes the normal elliptic integral of the second kind
and $\mbox{am}\,[u,k]$ is the amplitude, see \cite{bf}.
Using \cite[{\bf 113.02}, {\bf 122.06}]{bf},  we obtain
\[E(\mbox{am}\,[K_24K(k),k],k)=4K_2E(k)\]
where $E(k)$ is the complete elliptic integral of the second kind.
Thus,
\[
I_3(q)= 16m_2K_2(E(k)-(1-k^2)K(k)).
\]
Note that
for the linear static motor characteristic $q\mapsto I_3(q)$ is  constant.
Moreover,
\[
I_3^*:=\frac{1}{m_2}I_3(q)=\int_0^{K_12\pi/\Omega}\dot \theta^2\,dt>0.
\]
\newtheorem{id}[theorem]{Identity}
For the integrals $I_1$ and $I_2$ we have the following identity.
\begin{id}
\[\frac{3}{2}\Omega I_1(q)=(\cos{a}-\Omega^2)I_2(q).\]
\end {id}
\proof
Define $\eta:=K_12\pi/\Omega$ and compute:
\begin{eqnarray*}
I_1(q)&=& \int_0^\eta 2(\cos{\theta}-\cos{a})\cos{\theta}\sin{\Omega t}\,dt \\
&=& 2\int_0^\eta \cos^2{\theta}\sin{\Omega t}\,dt
- 2\cos{a}\int_0^\eta \cos{a}\sin{\Omega t}\,dt. \\
I_2(q)&=& -\frac{1}{\Omega}\int_0^\eta (\ddot \theta\sin{\theta}
    +\dot\theta^2\cos{\theta})\sin{\Omega t}\,dt \\
&=& -\frac{1}{\Omega}\int_0^\eta \ddot\theta\sin{\theta}\sin{\Omega t}\,dt
   -\frac{1}{\Omega}I_1(q) \\
&=& \frac{1}{\Omega}\int_0^\eta \sin^2{\theta}\sin{\Omega t}\,dt
   -\frac{1}{\Omega}I_1(q) \\
&=& -\frac{1}{\Omega}\int_0^\eta \cos^2{\theta}\sin{\Omega t}\,dt
   -\frac{1}{\Omega}I_1(q) \\
&=& -\frac{1}{\Omega}\left(
     \frac{1}{2}I_1(q)+\cos{a}\int_0^\eta \cos{\theta}\sin{\Omega t}\,dt
                     \right)
     -\frac{1}{\Omega}I_1(q) \\
&=& -\frac{3}{2\Omega}I_1(q)
-\frac{\cos{a}}{\Omega}\left(
   -\int_0^\eta(-\dot \theta\sin{\theta})(-\frac{1}{\Omega}\cos{\Omega t})\,dt
                       \right)\\
&=& -\frac{3}{2\Omega}I_1(q)+\frac{\cos{a}}{\Omega^2}I_2(q).
\,\mbox{\endproof}
\end{eqnarray*}
Also, with the definition
\[I_c:=\int_0^{K_12\pi/\Omega}\cos{\theta(t)}\cos\Omega t\,dt\]
we have a second identity.
\begin{id}
\[I_2(q)=\Omega I_c\sin{\Omega q}.\]
\end{id}
\proof
Define $\eta:=K_12\pi/\Omega$ and compute:
\begin{eqnarray*}
I_2(q)&=&-\int_0^\eta\frac{d}{dt}(\cos{\theta}(t+q))\cos{\Omega t}\,dt \\
&=&-\Omega\int_0^\eta\cos{\theta}(t+q)\sin{\Omega t}\,dt \\
&=&-\Omega\int_0^\eta\cos{\theta}(t)\sin{\Omega (t-q)}\,dt \\
&=&-\Omega\cos{\Omega q}\int_0^\eta\cos{\theta}(t)\sin{\Omega t}\,dt
+\Omega\sin{\Omega q}\int_0^\eta\cos{\theta}(t)\cos{\Omega t}\,dt.
\end{eqnarray*}
Since $t\mapsto \cos{\theta(t)}$ is an even function
\[
\int_0^\eta\cos{\theta}(t)\sin{\Omega t}\,dt=0.
\mbox{\endproof}
\]

Using the identities just obtained we have
\[
{\cal B}(p,q)=\left(
K_1\pi\lambda \,\mbox{e}^{-2p}-\frac{2}{3}
AI_c\,\mbox{e}^{-p}(\cos{a}-\Omega^2)
\sin{\Omega q},\;
-I_3+BI_c\Omega \,\mbox{e}^{-p}\sin{\Omega q}
             \right).
\]
Thus, $(p,q)$ is a zero of the bifurcation function if and only if this ordered
pair is a solution of the bifurcation equations
\begin{eqnarray*}
K_1\pi\lambda-\frac{2}{3}AI_c(\cos{a}-\Omega^2)\,\mbox{e}^p\sin{\Omega
q}&=&0,\\
I_3-BI_c\Omega e^{-p}\sin{\Omega q}&=&0.
\end{eqnarray*}
Such a zero is simple provided
\[\frac{4}{3}ABI_c^2(\cos{a}-\Omega^2)\sin{\Omega q}\cos{\Omega q}\ne 0.\]
To show the bifurcation is nondegenerate we must show
$I_c\ne 0.$ It turns out that the validity of this condition
depends on the resonance. This is the content
of the following proposition.
\begin{proposition}\label{prop:nonrot}
If $K_1$ and $K_2$ are relatively prime positive integers such that
$K_12\pi/\Omega=K_2 4K(k),$ then for $I_c$ to be nonvanishing it is
necessary and sufficient that $K_2=1$ and $K_1=2n$ for some positive
integer $n.$ In case this condition holds
\[
I_c=-2\left(\frac{\pi^2K_1}{k^2K(k)}\right)\frac{{\bf q}^{K_1/2}}{1-{\bf
q}^{K_1}}
=-4\pi K_2\Omega\frac{{\bf q}^{K_1/2}}{k^2(1-{\bf q}^{K_1})}
\]
where ${\bf q}:={\em e}^{-\pi K'/K}$ is Jacobi's {\sl nome}, {\em\cite[{\bf
1050}]{bf}}.
\end{proposition}
\proof
The proposition follows from the Fourier series representation of
$u\mapsto \mbox{sn}^2(u)$ given by
\[
(kK)^2\,\mbox{sn}^2(u)=
K^2-KE-2\pi^2\sum_{n=1}^\infty\frac{n{\bf q}^n}{1-{\bf q}^{2n}}\cos{2nx}
\]
where $x:=\pi u/(2K).$  (This formula is stated
without proof in \cite[p. 520]{ww}. A second Fourier series
expansion in \cite[{\bf 911.01}]{bf} seems to be incorrect. Thus, even though
a reference for the formula exists, we will
verify this series representation below.)
Define $\eta:=K_12\pi/\Omega.$ To prove the proposition, compute
\begin{eqnarray*}
I_c&=&-2\left(\frac{\pi}{kK}\right)^2\sum_{n=1}^\infty\frac{n{\bf q}^n}{1-{\bf
q}^{2n}}
\int_0^\eta\cos\left({\frac{n\pi}{K}u}\right)\cos{\Omega u}\,du \\
&=& -2\left(\frac{\pi}{kK}\right)^2\sum_{n=1}^\infty\frac{n{\bf q}^n}{1-{\bf
q}^{2n}}
\int_0^\eta\cos{\left(2n\frac{K_2}{K_1}\Omega u\right)}\cos{\Omega u}\,du.
\end{eqnarray*}
After the substitution $v:=\Omega u/K_1,$ we obtain
\begin{eqnarray*}
I_c&=& -2\left(\frac{\pi}{kK}\right)^2\sum_{n=1}^\infty\frac{n{\bf q}^n}{1-{\bf
q}^{2n}}
\frac{K_1}{\Omega}\int_0^{2\pi}\cos{2nK_2v}\cos{K_1 v}\,dv.
\end{eqnarray*}
Thus, $I_c$ vanishes unless $2nK_2=K_1.$ In particular,
$K_1$ must be even and $K_2$ must be a factor of $K_1.$ Since $K_1$ and $K_2$
are
relatively prime, $K_2=1.$ If $I_c\ne 0,$ then
\begin{eqnarray*}
I_c&=&-2\left(\frac{\pi}{kK}\right)^2\left(\frac{K_1}{\Omega}\right)
\left(\frac{K_1}{2}\right)\frac{{\bf q}^{K_1/2}}{1-{\bf q}^{K_1}}\pi   \\
&=&-2\frac{\pi^2K_1}{k^2 K}\frac{{\bf q}^{K_1/2}}{1-{\bf q}^{K_1}}
\end{eqnarray*}
as required.

To verify the Fourier series expansion we compute the value of
\[J:=\int_{-\pi}^{\pi} \,\mbox{sn}^2\left(\frac{2K}{\pi}x\right)
\,\mbox{e}^{imx}\,dx,
\hspace{.4in} m\ne 0\]
by contour integration around the parallelogram in the complex plane with
vertices $-\pi,$ $\pi,$ $\pi \tau$ and $\pi\tau-2\pi$ where $\tau:= iK'/K,$ cf.
\cite[p. 510]{ww}.
Using the fact that $u\mapsto \mbox{sn}(u)$ is doubly periodic with periods
$4K$ and $2iK'$ and $x\mapsto \mbox{e}^{imx}$ is periodic with period $2\pi,$
the path integrals along the edges of the parallelogram given by
$[\pi,\pi\tau]$
and $[\pi\tau-2\pi,-\pi]$ cancel. Also, an easy computation shows the integral
along
the edge $[\pi\tau,\pi\tau-2\pi]$ is
$-\mbox{e}^{im\pi\tau}\,\mbox{e}^{im\pi}J.$
Thus,
\[\left(1-\mbox{e}^{im\pi\tau}\,\mbox{e}^{im\pi}\right)J=2\pi i
\sum(\mbox{residues}).\]
The poles of $u\mapsto \mbox{sn}(u)$ reside at the points in the complex plane
congruent to $iK'$ and $2K+iK'$ modulo the periods of $\mbox{sn}.$
It follows that $\mbox{sn}^2(2Kx/\pi)\,\mbox{e}^{imx}$ has exactly two poles in
the
parallelogram. These poles are at the points $\pi\tau/2$ and $\pi\tau/2-\pi.$
To compute the residues, start with the Maclaurin series for $u\mapsto
\mbox{sn}(u)$
given by
\[\mbox{sn}(u)=u+O(u^3)\]
and the identity
\[\mbox{sn}(u+iK')=\frac{1}{k\,\mbox{sn}(u)}\]
to obtain
\[\mbox{sn}(u+iK')=\frac{1}{ku}+O(u).\]
Set $u+iK'=2Kx/\pi$ to get
\[\mbox{sn}(\frac{2K}{\pi}x)=\frac{\pi}{2kK(x-\pi\tau/2)}+O(x-\pi\tau/2)\]
and compute
\[
\mbox{sn}^2(\frac{2K}{\pi}x)\,\mbox{e}^{imx}=
\left(\frac{\pi}{2kK}\right)^2\,\frac{\mbox{e}^{im\pi\tau/2}}
{(x-\pi\tau/2)^2}
+\left(\frac{\pi}{2kK}\right)^2\frac{im\,\mbox{e}^{im\pi\tau/2}}
{(x-\pi\tau/2)}+O(1).
\]
Thus, the residue at $\pi\tau/2$ is
\[
\left(\frac{\pi}{2kK}\right)^2im\,\mbox{e}^{im\pi\tau/2}=\left(\frac{\pi}{2kK}\right)^2im{\bf q}^{m/2}.\]
Use the identity
\[ \mbox{sn}(u-2K+iK')=-\mbox{sn}(u+iK')\]
and a similar computation to compute the residue at $-\pi+\pi\tau/2.$
We find this residue to be
\[\left(\frac{\pi}{2kK}\right)^2im\,\mbox{e}^{-im\pi}{\bf q}^{m/2}.\]
{}From this it follows that
\[
J=-\frac{\pi^3}{2(kK)^2}\frac{m{\bf q}^{m/2}}{1-{\bf
q}^m\,\mbox{e}^{im\pi}}\left(1+\mbox{e}^{-im\pi}\right).
\]
Thus, the Fourier coefficient corresponding to $J$ vanishes unless $m=2n$ in
which
case
\[ J=-2\frac{\pi^3}{(kK)^2}\frac{n{\bf q}^n}{1-{\bf q}^{2n}}.\]
Since $x\mapsto \mbox{sn}^2(2Kx/\pi)$ is even, its Fourier series is a cosine
series. In fact, the Fourier coefficient of $\cos{2nx}$ is the real part of
$J/\pi.$
Since $J$ is real,
\[
\mbox{sn}^2(\frac{2K}{\pi}x)=a_0-2\left(\frac{\pi}{kK}\right)^2\sum_{n=1}^\infty
\frac{n{\bf q}^n}{1-{\bf q}^{2n}}\cos{2nx}.
\]
The constant term $a_0$ can be shown to agree with the stated  formula. But,
since we do not require its value here, the proof is left to the reader.
\endproof

By the proposition
we see there are (under appropriate choices of the constant
parameters)  bifurcating families of periodic solutions
for the full model equations at each nonrotational periodic motion,
of the gravitationally influenced rotor,  whose  period is an even multiple of
the
natural period of the support oscillator. In fact, if we impose the
nondegeneracy
conditions $K_1=2n$ and $K_2=1,$ then,
by eliminating $\sin{\Omega q}$ from the bifurcation equations, we find
\[
\mbox{e}^{2p}=\left(\frac{3\pi\lambda B\Omega}{A}\right)
\frac{n}{m_2 I_3^*(\cos{a}-\Omega^2)}.
\]
Thus, we can solve for $p$ provided $m_2(\cos{a}-\Omega^2)>0.$
Assuming this condition is satisfied and inserting $\mbox{e}^p$ into
the second bifurcation equation
we find there are two solutions for $q$ provided
$-1<\Delta<1$ for
\[
\Delta:=\left(\frac{3n\pi\lambda}{AB\Omega}\right)^{1/2}
  \left(\frac{I_3^*}{m_2(\cos{a}-\Omega^2)}\right)^{1/2}
  \left(\frac{m_2}{I_c}\right).
\]

The question arises as to how many resonant periodic solutions of the
unperturbed mathematical pendulum correspond to nondegenerate bifurcation
points.
It is clearly possible to obtain any preassigned finite number of simultaneous
bifurcations. However, it is not possible to have infinitely many.
To have infinitely many bifurcation points for a fixed set of parameter values
it is necessary that $\Delta$ remain bounded in the unit interval for
infinitely many integers $n$ such that $n=\Omega K(k)/\pi.$ To show
this is not the case, note $k\to 1$ as $n\to \infty,$ and use the
computations made above for $I_3^*$ and $I_c$ together with the fact that
$\cos{a}=1-2k^2$ to compute
\[
\Delta=-\sqrt{n}\,m_2k^2
\left(\frac{3\lambda}{16AB\Omega^3\pi}\right)^{1/2}
\left(\frac{1-{\bf q}^{2n}}{{\bf q}^n}\right)
\left(\frac{I_3^*(k)}{m_2(1-2k^2-\Omega^2)}\right)^{1/2}.
\]
We claim $\Delta$ grows without bound as
$k\to 1.$ Note first that as $k\to 1,$
\[
{\bf q}^n=\mbox{e}^{-\Omega K(\sqrt{1-k^2})}\to \mbox{e}^{-\Omega\pi/2}
\]
so the term $(1-{\bf q}^{2n})/({\bf q}^n)$ remains bounded.
Also, as $k\to 1$ we have  $K(k)-\ln(4/\sqrt{1-k^2})\to 0.$
Using these facts and the expression for $I_3^*(k)$ it follows that
$I_3^*(k)$ remains bounded as $k\to 1.$
Thus, all terms except the $\sqrt{n}$ term remain bounded. It follows that
$\Delta\to \infty$ as $k\to 1$ and $n\to \infty.$
However,  the fact that infinitely many different resonances
can lead to nondegenerate first order bifurcation to periodic solutions is
in marked contrast to the case
when the gravitational forces are considered small and the only nondegenerate
bifurcation occurs for the primary resonance.

To analyze the rotational motion of the rotor,  recall the mathematical
pendulum system defined on the phase plane is given by
\[\dot\theta=v,\hspace{.5in} \dot v=-\sin {\theta}+\epsilon h.\]
The rotational motions are naturally defined on the phase
cylinder that is obtained from the phase plane by
viewing the variable $\theta$ modulo $2\pi.$
There are two families of periodic solutions corresponding to
$\dot \theta <0$ and $\dot \theta >0.$ For definiteness we will treat the case
$\dot \theta <0,$ the other case is similar. In particular,
since we have changed the coordinates of the model equation
by $\theta\to -\theta-\pi,$ a positive rotation in the original
model equations
corresponds to a negative rotation here.
It is convenient to choose the (symplectic)
coordinate chart on the phase cylinder
given by the transformations
$x=\sqrt{-v}\cos{\theta},\,$ $y=\sqrt{-v}\sin{\theta}.$ The chart for the
second
case would be $x=\sqrt{v}\sin{\theta},\,$ $y=\sqrt{v}\cos{\theta}.$
This choice of coordinates ensures the divergence of the transformed vector
field vanishes and that the function $\beta_2(t,p)$ defined in \S 3 is zero.
In the $(x,y)$-plane the phase plane
system becomes
\begin{eqnarray*}
\dot x&=&y(x^2+y^2) +\frac{1}{2}xy(x^2+y^2)^{-3/2}
-\frac{\epsilon}{2}\frac{x}{x^2+y^2}h,\\
\dot y&=&-x(x^2+y^2) +\frac{1}{2}y^2(x^2+y^2)^{-3/2}
-\frac{\epsilon}{2}\frac{y}{x^2+y^2}h,\\
\end{eqnarray*}
where
\begin{eqnarray*}
h&=&-\left(M(\dot\theta)-Bz\sin \theta\right)=
-\left(M(-(x^2+y^2))-B\frac{yz}{\sqrt{x^2+y^2}}\right).
\end{eqnarray*}
We study
the above system  coupled as before to the support oscillator given by
\[\dot z=-\Omega w,\hspace{.5in} \dot w=\Omega z-\frac{\epsilon}{\Omega}g\]
where
\begin{eqnarray*}
g&=&-\left(F(z)+\lambda\dot z+A\dot \theta^2\cos\theta\right) \\
&=&-\left(F(z)-\lambda\Omega w+Ax(x^2+y^2)^{3/2}\right).
\end{eqnarray*}

Since the rotational motions correspond to curves in the phase plane which
do not intersect the $\theta$ axis, it is convenient to consider the $v$ axis
as a section for the flow.
On the trajectory passing through the point in the phase plane with coordinates
$(0,b),\;$ $|b|>2$  the first integral $I:=v^2/2-\cos{\theta}$
has the constant value  $I\equiv b^2/2-1.$ The case $\dot \theta <0$
corresponds
to $b<2.$
\[\frac{1}{2}\dot \theta^2=\cos{\theta}+\frac{1}{2}b^2-1.\]
Define $\varphi:=\theta/2$ and $k:=2/|b|$
to obtain equivalently
\[\dot \varphi^2=\frac{1}{k^2}\left(1-k^2\sin^2{\varphi}\right)\]
so that
\[
\int_0^{\varphi(t)}\frac{1}{\sqrt{1-k^2\sin^2{s}}}\,ds=\mbox{sgn(b)}\frac{t}{k}.
\]
Then, in terms of Jacobian elliptic functions
\[
\begin{array}{rclrcl}
\varphi(t)&=&\mbox{sgn(b)}\mbox{am}[t/k,k],\hspace{.5in}&
\dot \varphi(t)&=&\mbox{sgn(b)}\mbox{dn}[t/k,k]/k, \\
\cos{\varphi(t)}&=&\mbox{cn}[t/k,k],\hspace{.5in}&
\sin{\varphi(t)}&=&\mbox{sgn(b)}\mbox{sn}[t/k,k].
\end{array}
\]
Or, using the trigonometric double angle formulas,
\[
\begin{array}{rclrcl}
\theta(t)&=&2\mbox{sgn(b)}\,\mbox{am}[t/k,k],\hspace{.5in}&
\dot\theta(t)&=&2\mbox{sgn(b)}\,\mbox{dn}[t/k,k]/k, \\
\cos{\theta(t)}&=&1-2\,\mbox{sn}^2[t/k,k],\hspace{.5in}&
\sin{\theta(t)}&=&2\mbox{sgn(b)}\,\mbox{sn}[t/k,k]\,\mbox{cn}[t/k,k].
\end{array}
\]
Using these and the definition of the phase cylinder we have
\[
x(t)=\mbox{sgn(b)}\frac{2}{k}\,\mbox{dn}[t/k,k]\left(1-2\,\mbox{sn}^2[t/k,k]\right),
\hspace{.3in}
y(t)=
\frac{2}{k}\,\mbox{dn}[t/k,k]\,\mbox{sn}[t/k,k]\,\mbox{cn}[t/k,k].
\]
Also, observe the period $T$ of the periodic solution on the phase cylinder
with
initial value $(0,b)$ is given by $\theta(T/2)=\mbox{sgn(b)}\pi.$
Thus,
\[T=2k\,\mbox{am}^{-1}[\pi/2,k]=2kK(k)\]
and the resonance relation is given by
\[K_1\frac{2\pi}{\Omega}=K_22kK(k).\]

{}From \S 3,
\[
{\cal B}=\left(
\int_0^{K_12\pi/\Omega}wg\,dt,\hspace{.1in}
-\frac{1}{2}\int_0^{K_12\pi/\Omega}(x^2+y^2)h\,dt
         \right).
\]
It is preferable initially to express the components of ${\cal B}$ in phase
plane
coordinates:
\begin{eqnarray*}
\int_0^{K_12\pi/\Omega}wg\,dt&=&
  K_1\pi\lambda\,\mbox{e}^{-2p}-A\,\mbox{e}^{-p}I_1^r(q),\\
-\frac{1}{2}\int_0^{K_12\pi/\Omega}(x^2+y^2)h\,dt&=&
-\frac{1}{2}I_3^r(q)-\frac{1}{2}B\,\mbox{e}^{-p}I_2^r(q),
\end{eqnarray*}
where
\begin{eqnarray*}
I_1^r(q)&:=&
\int_0^{K_12\pi/\Omega}(\dot\theta(t+q))^2\cos{\theta(t+q)}\sin{\Omega
t}\,dt,\\
I_2^r(q)&:=&\int_0^{K_12\pi/\Omega}\dot \theta(t+q)\sin{\theta(t+q)}\cos{\Omega
t}\,dt, \\
I_3^r(q)&:=&\int_0^{K_12\pi/\Omega}\dot \theta(t+q)M(\dot\theta(t+q))\,dt. \\
\end{eqnarray*}
Also, we define
\begin{eqnarray*}
I_c^r(q)&:=&\int_0^{K_12\pi/\Omega}\cos{\theta(t)}\sin{\Omega t}\,dt.
\end{eqnarray*}
As in the case of the nonrotational motions we have the following identities:
\begin{id}
\begin{eqnarray*}
\frac{3}{2}\Omega I_1^r(q)&=&-\left(\frac{2}{k^2}-1+\Omega^2\right)I_2^r(q), \\
I_2^r(q)&=&\Omega I_c^r\sin{\Omega q}. \\
\end{eqnarray*}
\end{id}
Using these identities, we find
\[{\cal B}(p,q):=({\cal B}_1(p,q),{\cal B}_1(p,q))\]
where
\begin{eqnarray*}
{\cal B}_1(p,q)&=& K_1\pi\lambda\,\mbox{e}^{-2p}+
\frac{2}{3k^2}A\,\mbox{e}^{-p}\left(2+k^2(\Omega^2-1)\right)I_c^r\sin{\Omega
q},\\
{\cal B}_2(p,q)&=&I_3^r-\frac{1}{2}B\,\mbox{e}^{-p}\Omega I_c^r\sin{\Omega q}.
\end{eqnarray*}
\begin{proposition}
If $K_1$ and $K_2$ are relatively prime positive integers such that
$K_12\pi/\Omega=K_22kK(k),$ then for $I_c^r$ to be nonvanishing it is
necessary and sufficient that $K_2=1.$
In case this condition holds
\begin{eqnarray*}
I_c^r&=&4\left(\frac{\pi^2K_1}{kK(k)}\right)\frac{{\bf q}^{K_1}}{1-{\bf
q}^{2K_1}}
=4\pi\Omega\frac{{\bf q}^{K_1}}{1-{\bf q}^{2K_1}} \\
\end{eqnarray*}
where ${\bf q}:={\em e}^{-\pi K'/K}$ is Jacobi's {\sl nome}.
\end{proposition}
\proof
The integral $I_c^r$  is computed as in proposition \ref{prop:nonrot} using the
Fourier series for $\,\mbox{sn}^2(u).$ In fact,
\begin{eqnarray*}
I_c^r&=& -2\int_0^{K_12\pi/\Omega}\,\mbox{sn}^2(t/k)\cos{\Omega t}\,dt \\
&=&4\left(\frac{\pi}{kK}\right)^2\sum_{n=1}^\infty\frac{n{\bf q}^n}{1-{\bf
q}^{2n}}
\int_0^{K_12\pi/\Omega}\cos\left(\frac{n\pi}{kK}t\right)\cos{\Omega t}\,dt. \\
\end{eqnarray*}
After the change of variables $v:=\Omega t/K_1$ and substitution from the
resonance relation we obtain
\begin{eqnarray*}
I_c^r&=&4\left(\frac{\pi}{kK}\right)^2\frac{kKK_2}{\pi}
\sum_{n=1}^\infty\frac{n{\bf q}^n}{1-{\bf q}^{2n}}
\int_0^{2\pi}\cos{nK_2v}\cos{K_1v}\,dv.
\end{eqnarray*}
Thus, $I_c^r$ vanishes unless $nK_2=K_1.$ Since $K_1$ and $K_2$ are relatively
prime, this means $I_c^r\ne 0$ exactly when $K_2=1$ and $K_1$ is arbitrary.
In this case we obtain
\[
I_c^r=4\left(\frac{\pi^2K_1}{kK(k)}\right)\frac{{\bf q}^{K_1}}{1-{\bf
q}^{2K_1}}.
\]
\endproof

Finally, we compute $I_3^r$ under the assumption $M(\dot
\theta)=m_1+m_2\dot\theta.$
For this we have
\[I_3^r=m_1\int_0^{K_12\pi/\Omega}\dot\theta(t+q)\,dt
+m_2\int_0^{K_12\pi/\Omega}\dot\theta^2(t+q)\,dt.
\]
In the present case we find, using the resonance relation  and
the periodicity,
\[\int_0^{K_12\pi/\Omega}\dot\theta(t+q)\,dt=-2\pi K_2.\]
Also, as before,
\begin{eqnarray*}
\int_0^{K_12\pi/\Omega}\dot\theta^2(t+q)\,dt&=&
\int_0^{K_12\pi/\Omega}(2\cos(\theta(t+q))+b^2-2)\,dt \\
&=&(b^2-2)K_1\frac{2\pi}{\Omega}
+2\int_0^{K_12\pi/\Omega}1-2\mbox{sn}^2[t/k,k]\,dt \\
&=& \frac{4}{k}E(\mbox{am}\,[2K_2K(k),k],k)\\
&=& 8K_2\frac{E(k)}{k}
\end{eqnarray*}
Thus, we have
\[I_3^r(q)=m_1\pi K_2-m_24K_2\frac{E(k)}{k}.\]

By the proposition
we see there are (under appropriate choices of the constant
parameters)  bifurcating families of periodic solutions
for the full model equations at each rotational periodic motion,
of the gravitationally influenced rotor, whose  period is an integer
multiple of the natural period of the support oscillator.
The fact that the resonances are not restricted to {\em even} multiples of the
period of the support oscillator as in the case of nonrotational motions
is perhaps expected since near the separatrix between rotational and
nonrotational motions the nonrotational
periods are twice as long as the rotational periods.
More precisely, if we impose the nondegeneracy
condition $K_2=1,$ the bifurcation points are the simple solutions of the
equations
\begin{eqnarray*}
\lambda K_1\pi e^{-2p}+\frac{2A}{3k^2}(2+k^2(\Omega^2-1))4\pi\Omega
\frac{{\bf q}^{K_1}}{1-{\bf q}^{2K_1}}e^{-p}\sin{\Omega q}&=&0,\\
m_1\pi-m_24\frac{E(k)}{k}+2\pi B\Omega^2
\frac{{\bf q}^{K_1}}{1-{\bf q}^{2K_1}}e^{-p}\sin{\Omega q}&=&0.
\end{eqnarray*}
By eliminating $\sin{\Omega q}$ from these equations, we find
\[
\mbox{e}^{2p}=\Delta:=\frac{3\pi\lambda B\Omega K_1k^3}
{4A(m_1\pi k-m_24E(k))(2+k^2(\Omega^2-1))}.
\]
Thus, we can solve for $p$ provided $(m_1\pi k-m_24E(k))(2+k^2(\Omega^2-1))>0.$
Assuming this condition is satisfied and inserting $\mbox{e}^p$ into
the second bifurcation equation
we find $\sin{\Omega q}:=\Lambda$ where
\[\Lambda=
-\left(\frac{3\pi\lambda B\Omega K_1 k^3}
{4A(m_1\pi k-m_24E(k))(2+k^2(\Omega^2-1))}\right)^{1/2}
\left(\frac{m_1\pi k-m_24E(k)}{2\pi B\Omega^2k}\right)
\frac{1-{\bf q}^{2K_1}}{{\bf q}^{K_1}}.
\]
Thus, there are two solutions for $q$ provided
$-1<\Lambda<1.$
In addition, it is easy to compute the Jacobian of the
two bifurcation equations and deduce that the solutions of the
bifurcation equations  will
both be simple provided $\cos{\Omega q}\ne 0.$ This is as it should be since
the solutions are simple when there are two values of $q$ and not simple
at the bifurcation points given by $\sin{\Omega q}=\pm 1.$

As in the case of the nonrotational motions, if the parameters are fixed, then
there are only finitely many resonant motions of the rotor for which
the condition $-1<\Lambda<1$ is satisfied.
This follows as before by showing  $\Lambda$ is unbounded as $k\to 1$ and
$K_1\to \infty.$
Thus, again for rotational motions under a strong gravitational force,
infinitely many resonant solutions can lead to nondegenerate first order
bifurcation, but only a finite number of these are excited for a fixed set
of parameter values.

We end this section with a useful observation. The divergence
of the perturbed vector field, computed in $(z,w,x,y)$-coordinates, is
constant. In fact, the divergence is simply $-\epsilon(\lambda+m_2).$
This is reasonable since $\lambda$ and $m_2$ are coefficients of damping
in the system. Abel's formula applied to the linear variational equations
as in \cite[p. 156]{wiggins2} implies the determinant of the linearized
Poincar\'e map is given by
\[\mbox{det}\,DP(\xi,\epsilon)=e^{-\epsilon(\lambda+m_2)K_12\pi/\Omega}.\]
Thus, the linearized Poincar\'e map contracts volume and the perturbed
periodic solutions found by our bifurcation method are all
saddles and sinks. In particular, this shows entrainment (capture)
is possible.

\subsection{Remarks, Experiments and Speculation}
We have just shown there exist choices of the parameters in our model equations
such that several periodic solutions, corresponding to rotational motions of
the rotor,
can coexistent. Moreover, these periodic solutions
in the four dimensional phase space are all
saddles or sinks. In order to determine the dynamics of the
system, we would like further stability information about these periodic
solutions.
Rigorous stability information may be obtained from
a second order bifurcation analysis.  However, we mention that
the bifurcating
families occur in pairs, corresponding to the solutions of the equation
$\sin\Omega q=\Lambda.$ It is likely that, generically, one bifurcating
family with consist of sinks  the other consists of saddles.
The basin of attraction of a periodic solution corresponding to a sink
is the region in phase space ``captured into resonance'' or, in other language,
it is the entrainment domain.
Of course, there is no obvious reason why such a periodic solution will be
globally attracting, thus solutions starting outside of the basin
of attraction have a different fate.
On the other hand, a saddle periodic solution may
have a
one, two or three dimensional stable manifold.
Solutions starting near the stable manifold
may remain near the saddle periodic solution on a very long time scale,
appearing
to be captured, only to eventually leave the vicinity of the saddle periodic
solution along its unstable manifold to
pass near a second saddle or perhaps become entrained
to a stable periodic solution.
If there are several such saddles, this behavior may
be very complex.

At the end of the last section we showed the linearized Poincar\'e  map
contracts
volume as a lemma to show the perturbed periodic solutions are saddles and
sinks.
In contrast to the similar  analysis of a single forced oscillator, e.g.
\cite[p. 157]{wiggins2} or \cite[p. 207]{gh}, we can not conclude there are no
invariant closed curves for the {\em three} dimensional perturbed Poincar\'e
map.
In other words periodic sinks may coexists with more complicated attractors.
Before discussing this possibility more fully we mention that
the analysis completed above only considers the bifurcation of periodic
solutions
from periodic solutions of the unperturbed oscillators at resonance. In the
example
with a strong gravitational force, the mathematical pendulum has, in the phase
cylinder, a hyperbolic saddle point corresponding to its unstable equilibrium
state, and this rest point has a pair of associated homoclinic
trajectories. The dynamics of the perturbed system near the corresponding
trajectories
in the four dimensional phase space of the coupled system can perhaps
be determined
to some extent by analyzing an appropriate ``Melnikov'' integral.
Such an analysis might show the presence of horseshoes.
In any case, the existence of complicated attractors remains to be established.
But, as an excursion in this direction, we have considered a decoupled
specialization
of our model equations in order to obtain a two dimensional Poincar\'e map and
the possibility of  visual representations of some aspects of the dynamics.
For this we consider the system
\begin{eqnarray*}
\ddot z+\Omega^2 z&=&0, \\
\ddot \theta+\sin{\theta}&=&-\epsilon(m_1+m_2\dot\theta-Bz\sin{\theta}).
\end{eqnarray*}
It may be viewed as a single parametrically excited mathematical pendulum.
As before, to study the rotational motions, we consider (symplectic)
polar coordinates on the phase cylinder to obtain
\begin{eqnarray*}
\dot x&=&y(x^2+y^2) +\frac{1}{2}xy(x^2+y^2)^{-3/2}
-\frac{\epsilon}{2}\frac{x}{x^2+y^2}h,\\
\dot y&=&-x(x^2+y^2) +\frac{1}{2}y^2(x^2+y^2)^{-3/2}
-\frac{\epsilon}{2}\frac{y}{x^2+y^2}h,\\
\end{eqnarray*}
where
\begin{eqnarray*}
h= -m_1+m_2(x^2+y^2)+B\frac{y}{\sqrt{x^2+y^2}}\mbox{e}^{-p}\cos{\Omega t}.
\end{eqnarray*}
A comparison of the analysis for the coupled system with the analysis
for the single ``forced'' oscillator as presented in \cite{cccfe,cccc}
shows we have already computed the bifurcation function for this system.
Here, $p$ is just a parameter, and the scalar bifurcation function
is just ${\cal B}_2(q)$ as computed above. In fact, for the $(K_1:K_2)$
resonance,
the bifurcation equation
is
\[
{\cal B}_2(q)=
\left\{
\begin{array}{lr}
m_1\pi-m_24\frac{E(k)}{k}+2\pi B\Omega^2
\frac{{\bf q}^{K_1}}{1-{\bf q}^{2K_1}}e^{-p}\sin{\Omega q}=0,&\hspace{.2in}
{\rm if}\; K_2=1 \\
K_2(m_1\pi-m_24\frac{E(k)}{k}),&\hspace{.2in}
{\rm if}\; K_2\ne 1.
\end{array}
\right.
\]
In case $K_2=1,$  the bifurcation equation ${\cal B}_2(q)=0$ is
equivalent to $\sin{\Omega q}:=\Lambda$ where
\[\Lambda=
-\mbox{e}^p\left(\frac{m_1\pi k-m_24E(k)}{2\pi B\Omega^2k}\right)
\frac{1-{\bf q}^{2K_1}}{{\bf q}^{K_1}}.
\]
Thus, there are two solutions for $q$ provided
$-1<\Lambda<1.$
Here, the linearized Poincar\'e map is still area contracting
(the divergence is $-\epsilon m_2$) so the periodic solutions are
again saddles and sinks. However, even in this case we {\em can not} conclude
there are no invariant curves in the Poincar\'e section. This is
due to the fact
that, for the rotational motions, the system is defined on an annulus in the
phase cylinder whose inner boundary is the separatrix of the unperturbed
mathematical pendulum. This fact
is reflected in the singularity of the $(x,y)$-coordinates
at $x^2+y^2=-\dot \theta=0.$ In other words, in the $(x,y)$ section
the region corresponding to rotational motion is an annular region
surrounding the origin. More precisely, the unperturbed rotational
solutions correspond to solutions (outside the separatrices) with energies
\begin{eqnarray*}
E&=&\frac{1}{2}\dot\theta^2+1-\cos{\theta}\\
&=& \frac{1}{2}(x^2+y^2)^2+1-\frac{x}{\sqrt{x^2+y^2}}>1.
\end{eqnarray*}
When the system is perturbed, solutions can cross into the region with
$E<1 $ and
then eventually cross the curve $\dot \theta=0$ where the vector field is
singular. Thus, the area of the region corresponding to rotational motions is
not
preserved.  Of course, the fact that the linearized Poincar\'e map is area
contracting does imply there is at most one invariant curve. If there were two
invariant curves, the annular region bounded by these curves would be
invariant.
Numerical experiments suggest that in fact
invariant curves exit. This suggests a similar phenomenon is possible for the
coupled system. But, at present we do not know how to examine this possibility
rigorously.
\begin{figure}[t]
\vspace{3.5in}
\caption{Schematic representation of basin boundaries for attractors in the
perturbed Poincar\'e map for the rotational motions of the
parametrically excited mathematical pendulum.}
\end{figure}

We have investigated the dynamics of the uncoupled system in the region of
parameter space corresponding to parameter values where the {\em coupled}
system has
periodic solutions arising from the bifurcation theory given previously.
A useful example is provided by
the following choice of parameters:
\[K_1=1,\;\; \Omega =4,\;\; A=4,\; \lambda=0,\;\; B=4,\;\; m_1=10\]
with $m_2$ and $p$  variable.
Let $\Gamma$ denote the $(K_1:K_2)=(1:1)$  resonant periodic solution
of the mathematical pendulum. This solution is given by
the elliptic modulus $k_*\approx 0.47.$ Or, in the original coordinates, it
is the solution starting at $(\theta,\dot\theta)=(0,-2/k_*).$ Recall a
necessary
condition for the
bifurcation equations obtained for the {\em coupled}
system to have solutions in this case
is
$m_1\pi k_*-m_2 4 E(k_*)>0.$ Thus, for the given parameters, we must have
$0<m_2<5\pi k_*/(2E(k_*))\approx 2.495.$ If this condition is satisfied,
$p$ is determined by the formula $e^{2p}=\Delta$ given previously.
The unperturbed Poincar\'e map  for the uncoupled system giving the return
to the $(x,y)$-plane after time $2\pi/\Omega$ has $\Gamma$ as an invariant
curve. In fact, the unperturbed Poincar\'e map is the identity on $\Gamma.$
For small positive $\epsilon$ we have proved there are two fixed points
for the perturbed Poincar\'e map near the zeros of ${\cal B}_2(q)$ and that
these fixed points
correspond to the persistent
periodic solutions. For the resonances with $K_2\ne 1$ the bifurcation
function reduces to
$b(k):= K_2(m_1\pi-m_2 4E(k)/k)$ and is independent of $q.$
It is easy to see the dense set of resonant orbits such that
$K_1/K_2<1$ lie ``outside'' $\Gamma$ in the Poincar\'e section while
the dense set of resonant orbits such that $K_2\ne 1$ and
$K_1/K_2>1$ lies ``inside'' $\Gamma.$ Moreover, the reduced bifurcation
function $b$ is positive inside and on $\Gamma.$ That is, there is an
unperturbed
periodic solution $\Gamma_0$ of the mathematical pendulum corresponding
to some $k_0(m_2)<k_*$ such that $b(k_0)=0.$ In particular, $\Gamma_0$
surrounds $\Gamma,\,$ $b(k)<0$ for all
$k<k_0$ and the resonant orbits outside $\Gamma_0$ correspond
to resonances such that $K_1/K_2<1.$
Thus, for these resonant orbits $K_2\ne 1$ and ${\cal B}_2=b.$
In general,
some of the resonant orbits corresponding to $K_2=1$ and $K_1>1$
can be excited and additional periodic solutions can occur. But,
for our choice of parameters this does not happen. Thus,
in a similar manner to the
discussion in \cite[pp 161-175]{wiggins2}, we observe that perturbed
trajectories of the Poincar\'e map tend to drift out toward $\Gamma_0$ from the
region inside $\Gamma_0$ except for the resonance layer near $\Gamma$  and tend
to
drift in toward $\Gamma_0$ from the outside. In particular, there are no
periodic
solutions excited by the perturbation except for those on $\Gamma.$
The existence of a periodic sink and a periodic saddle for the perturbed
Poincar\'e map corresponding to
the perturbed periodic solutions is consistent with these facts. This is
exactly the
situation observed in numerical simulation. In addition, the
positions
of the bifurcation points as predicted independently by solving
the bifurcation equations $e^{2p}=\Delta$ and $\sin\Omega q=\Lambda$
is also confirmed.
However, the facts about the sign of $b$ and the implied drift
directions for the perturbed solutions indicate  there is also the
possibility of a non periodic attractor $\Gamma_\epsilon$ near $\Gamma_0$
coexistent with the periodic sink.
Our numerical experiments confirm the
existence of such an invariant attracting set. It
appears to be a smooth curve
for $0<m_2<2.459$ and for all sufficiently small $\epsilon>0.$
Of course, from the discussion above,
the periodic sink lies inside the region bounded by this invariant curve.
Because there are two attractors,
the entrainment domain (the basin of attraction
of the periodic sink) shares a common boundary with the basin of attraction
of the invariant curve. Fig.\ 1 shows, schematically, the basins of attraction
for
the two attractors. Aside from the fact that the spiral basin of attraction of
the
periodic sink is very thin for small $\epsilon,$ we also see that
solutions with
initial values ``outside''
$\Gamma_\epsilon$ are never entrained to the periodic
sink. We expect the entrainment domain for the coupled system to be
at least as complex.

\end{document}